\documentclass[prl,aps,twocolumn,superscriptaddress,balancelastpage]{revtex4-1}

\usepackage{latexsym,amssymb,bm,bbold,amsfonts,amstext,graphicx,bbm,bm,relsize,dsfont,times}
\usepackage[dvipsnames]{xcolor}
\usepackage{enumerate}
\usepackage{amsmath} 
\usepackage{amsthm} 
\usepackage[unicode=true, breaklinks=false, pdfborder={0 0 1}, backref=false, colorlinks=true, linkcolor=blue, citecolor=blue]{hyperref}
\usepackage{subfiles}
\usepackage{tcolorbox}
\usepackage{tikz}

\definecolor{ao}{rgb}{0.0, 0.5, 0.0}

\begin{document}

\def\be{\begin{equation}}
\def\ee{\end{equation}}
\def\bea{\begin{eqnarray}}
\def\eea{\end{eqnarray}}

\newcommand{\mmm}[1]{\textcolor{red}{#1}}

\title{Information flow and error scaling for fully-quantum control}

\author{Stefano Gherardini}
\affiliation{\mbox{Department of Physics and Astronomy \& LENS, University of Florence,} via G. Sansone 1, I-50019 Sesto Fiorentino, Italy.}
\affiliation{\mbox{Istituto Nazionale di Fisica Nucleare (INFN)}, Italy.}

\author{Matthias M. M\"{u}ller}
\affiliation{\mbox{Peter Gr\"{u}nberg Institute -- Quantum Control (PGI-8)}, Forschungszentrum J\"{u}lich, D-52428 J\"{u}lich, Germany.}

\author{Simone Montangero}
\affiliation{\mbox{Istituto Nazionale di Fisica Nucleare (INFN)}, Italy.}
\affiliation{\mbox{Department of Physics and Astronomy ``G. Galilei'', University of Padua}, I-35131 Italy.}

\author{Tommaso Calarco}
\affiliation{\mbox{Peter Gr\"{u}nberg Institute -- Quantum Control (PGI-8)}, Forschungszentrum J\"{u}lich GmbH, J\"{u}lich, Germany.}
\affiliation{\mbox{Institute for Theoretical Physics}, University of Cologne, D-50937 Cologne, Germany.}

\author{Filippo Caruso}
\affiliation{\mbox{Department of Physics and Astronomy \& LENS, University of Florence,} via G. Sansone 1, I-50019 Sesto Fiorentino, Italy.}

\begin{abstract}
The optimally designed control of quantum systems is playing an increasingly important role to engineer novel and more efficient quantum technologies. Here, in the scenario represented by controlling an arbitrary quantum system via the interaction with an another optimally initialized auxiliary quantum system, we show that the quantum channel capacity sets the scaling behaviour of the optimal control error. Specifically, we prove that the minimum control error is ensured by maximizing the quantum capacity of the channel mapping the initial control state into the target state of the controlled system, i.e., optimizing the quantum information flow from the controller to the system to be controlled. Analytical results, supported by numerical evidences, are provided when the systems and the controller are either qubits or single Bosonic modes and can be applied to a very large class of platforms for controllable quantum devices.
\begin{description}
\item[PACS numbers]03.67.-a, 02.30.Yy, 42.50.Dv
\end{description}
\end{abstract}

\maketitle

Quantum control theory studies the steering of a quantum system from an initial state to a desired target one, by means of a control system that can be either classical or quantum\,\cite{WarrenScience1993,SteckPRL2004,Edwards2005,DAlessandroBook2007,GordonPRL2008,NurdinAut2009,BrifNJP2010,WisemanBook,SayrinNature2011,AltafiniTAC2012,TicozziTAC2012,TicozziTAC2012_2,Glaser2015,Koch2016,GirolamiPRL2019,GherardiniBattery2019}. Quantum control has played a key role in recent quantum technology breakthroughs\,\cite{Taminiau14,RossiNature2018,AruteNature2019,OmranScience2019,SongScience2019}, and, thus, the problem of identifying a universal relation for the scaling of the control error with the relevant parameters of system and control knobs is no longer only academic but also practical and even decisive for the success of any quantum platform. This especially holds if the control action is provided by the interaction between a quantum system to be controlled and an auxiliary one, namely the quantum controller, then the control problem is denoted as \emph{coherent-quantum} or \emph{fully-quantum}\,\cite{LloysPRA2000,NelsonPRL2000,SekatskiQuantum2017}. 

In the scenario in which a quantum system is controlled by optimal coherent pulses, which can be engineered for instance via the Krotov method\,\cite{KonnovAiT1999}, the gradient ascent pulse engineering (GRAPE)\,\cite{KhanejaJMR2005} and the (dressed) chopped random basis (dCRAB) optimal control algorithms\,\cite{DoriaPRL2011,CanevaPRA2011,RosiPRA2013,HoyerNJP2014,RachPRA2015,LovecchioPRA2016,vanFrankSciRep2016}, the cost function (or landscape), which quantifies the error in performing the desired control task, may have many local minima. This can entail the ``entrapment'' of the optimization procedure and, thus, the impossibility of completing the control task\,\cite{BrifNJP2010,RabitzScience2004,RussellJPA2017,WuJMC2019}, especially in the open quantum systems case\,\cite{PetruccioneBook,Caruso_RevModPhys_2014,Koch2016}. However, the situation is different when the controller is another quantum system. In this case, indeed, the control landscape, provided by the error of the control task as a function of the input state of the quantum controller, is usually convex and the optimal solution can be straightforwardly found by optimization or analytic solutions, independently of the complexity in preparing the initial state of the quantum controller\,\cite{WuJMC2019}.

A similar statement could be made about the complexity of the control, identified by its information content. For the classical control of a quantum system, it was numerically found that the control complexity has to correspond at least to the dimension of the quantum system\,\cite{Moore2012,CanevaPRA2014,RachPRA2015}. This can be explained by considering the control problem as a (classical) communication channel between the control and the system\,\cite{LloydPRL2014,Mueller2020}, where the control pulse is interpreted as a communication signal whose correct reception means the complete attainment of the desired control task. Also in solving fully-quantum control problems, universal results from information and communication theories could be used. In this regard, it is well-known that any physical quantum process (thus, also a quantum system interacting with a quantum controller) can be generally represented as a quantum channel mapping an initial state to a final one\,\cite{Caruso_RevModPhys_2014}. 

\begin{figure}[t!]
\centering
\includegraphics[scale=0.32]{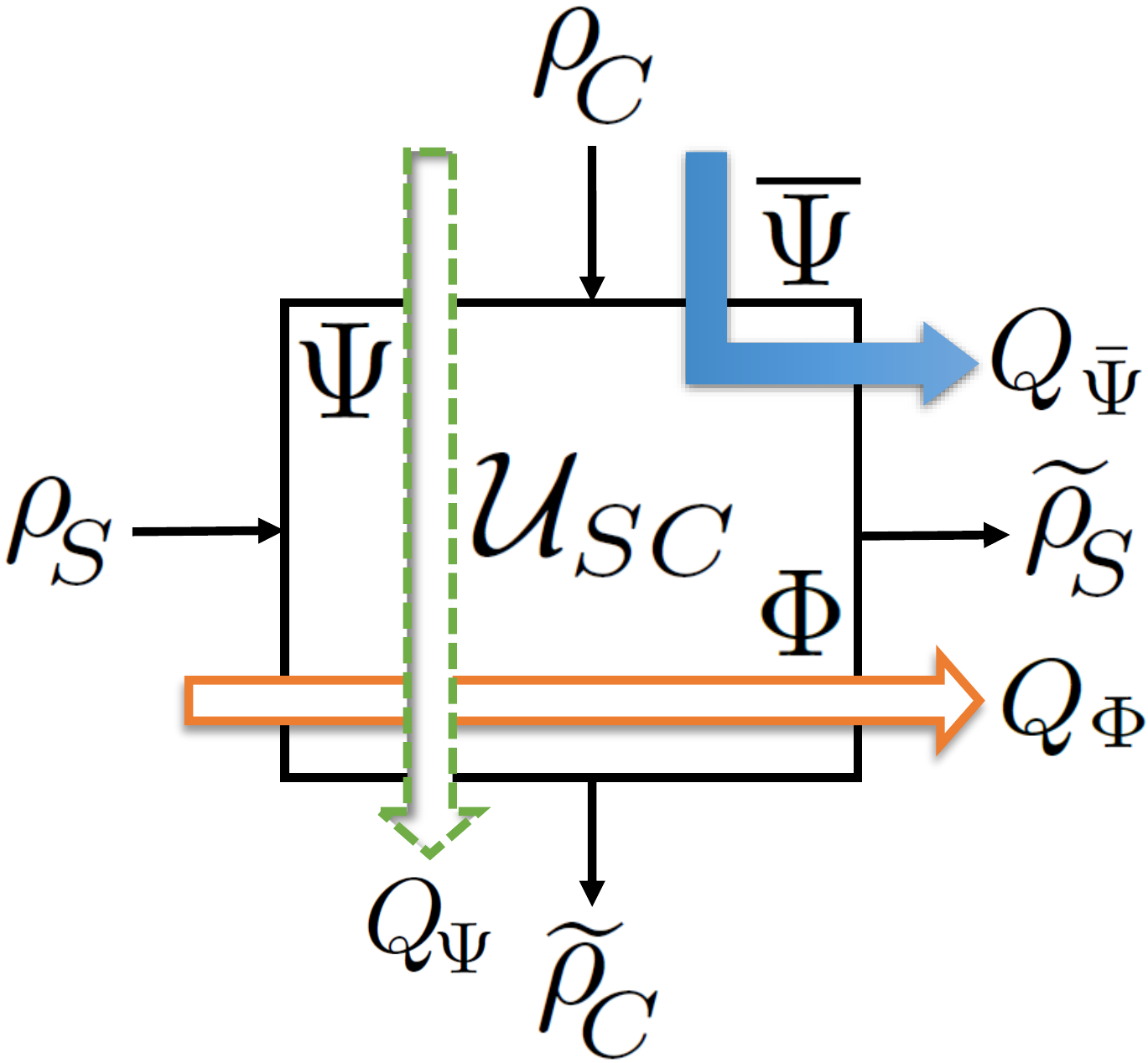}
\caption{Pictorial representation of a fully-quantum control procedure given by the unitary interaction $\mathcal{U}_{SC}$ between the quantum system $S$ to be controlled and the quantum controller $C$. The dashed green and solid orange arrows identify the channels $\Psi$ and $\Phi$, with quantum capacities $Q_{\Psi}$ and $Q_{\Phi}$, modelling the reduced dynamics of the quantum controller $C$ and quantum system $S$, which map $\rho_C$ and $\rho_S$ into $\widetilde{\rho}_{C}$ and $\widetilde{\rho}_{S}$ respectively. The blue arrow refers to the complementary channel $\overline{\Psi}$ with quantum capacity $Q_{\bar{\Psi}}$ responsible of the quantum control performance and mapping $\rho_C$ into $\widetilde{\rho}_{S}$.}
\label{fig:fig1}
\end{figure}

In this Letter, as depicted in Fig.\,\ref{fig:fig1}, we formalize the fully-quantum control problem according to the quantum channels language, commonly used in quantum information and communication theory. This allows us to determine the analytical expression of the control error scaling in reaching a desired target state. Specifically, we find that the control error scales exponentially with the quantum channel capacity of the channel linking the initial state of the controller with the output state of the controlled system (control channel). As a result, the performance of a fully-quantum control is exponentially enhanced as the quantum channel capacity of the control channel increases. We provide analytical results that show how the maximization of the control channel's quantum channel capacity decreases the control error, when in each configuration the control is optimized via the initial state of the quantum controller. As discussed in more details below, these results are expected to have remarkable implications in, among others, state preparation of many-body quantum systems\,\cite{OmranScience2019,SongScience2019,LovecchioPRA2016}, the realization of photonic links\,\cite{SimonNatPhys2007} between quantum processors, and long-distance communication through quantum carriers\,\cite{HammererRMP10}.

\paragraph{Quantum channels \& control problem.--}

Let us consider a bipartite quantum system composed of the system $S$ to be controlled and the auxiliary one $C$ representing the controller. The goal of the control is to bring $S$ from the initial density operator $\rho_{S}$ to the target $\hat{\rho}_S$ chosen by the user through a properly designed dynamical transformation. The latter and also the final state of the system depend on the initial (input) state $\rho_{C}$ of the controller $C$. Here, the control problem is to find the value of $\rho_{C}$ that minimizes the distance between the final and target states. To this end, the quantum controller $C$ has to be optimally initialized.

Any physical transformation performed on a quantum system can be generally described by a family of completely-positive trace-preserving (CPTP) maps $\Phi[\cdot]:\rho_{S}\rightarrow\widetilde{\rho}_{S}\equiv\Phi[\rho_{S}]$, with $\rho_{S}$ and $\widetilde{\rho}_{S}$ denoting respectively the initial and final density operator of $\mathcal{S}$ before and after the transformation. In Fig.\,\ref{fig:fig1} we show a pictorial scheme identifying the fully-quantum control problem given by the interaction between $S$ and the quantum controller $C$. The composite system $SC$ is initially prepared in the product state $\rho_{\rm in} = \rho_{S} \otimes \rho_{C}$, where $\rho_{C}$ is denoted as control state.

Under the assumption that the target state $\hat{\rho}_S$ belongs to the set of density operators that can be reached by the system\,\cite{footnote_1}, to solve the control problem we need to find the optimal value of $\rho_{C}$ allowing for the equality
\begin{equation}\label{eq:equality_rhoC}
\widetilde{\rho}_{S}=\Phi\left[\rho_{S}\right] 
\equiv {\rm Tr}_{C}\left[\mathcal{U}_{SC}\,\rho_{\rm in}\,\mathcal{U}_{SC}^{\dagger}\right] = \hat{\rho}_S 
\end{equation}
where $\mathcal{U}_{SC}$ is the unitary map describing the physical transformation of the composite system. If the target state cannot be reached by the system, the equality (\ref{eq:equality_rhoC}) does not have a solution and this unavoidably leads to a non zero error value $\varepsilon$. The control error $\varepsilon$ is commonly expressed as a function of the Uhlmann fidelity $\mathfrak{F}(\hat{\rho}_{S},\widetilde{\rho}_S) \geq 0$  (and $\leq 1$) between the target state $\hat{\rho}_{S}$ and the final state $\widetilde{\rho}_S$, namely 
$\varepsilon = 1 - \mathfrak{F}(\hat{\rho}_{S},\widetilde{\rho}_S)$, with $\mathfrak{F}(\hat{\rho}_{S},\widetilde{\rho}_S) \equiv {\rm Tr}\sqrt{\sqrt{\hat{\rho}_{S}}\widetilde{\rho}_S\sqrt{\hat{\rho}_{S}}}$ \cite{Uhlmann1976}. In this case, solving the control problem corresponds to finding the optimal state $\rho_{C}$ that minimizes the residual control error. Moreover, as illustrated in Fig.\,\ref{fig:fig1}, $\Psi[\rho_C] \equiv {\rm Tr}_{S}[\mathcal{U}_{SC}\,\rho_{\rm in}\,\mathcal{U}_{SC}^{\dagger}]$ defines the CPTP map transforming $\rho_{C}$ into $\widetilde{\rho}_{C}$, while the super-operator mapping $\rho_{C}$ into $\widetilde{\rho}_{S}$ is given by the \emph{complementary} quantum channel
$\overline{\Psi}[\rho]:\rho_{C}\rightarrow\widetilde{\rho}_{S}$. As we will show, in the quantum control problem represented in Fig.\,\ref{fig:fig1}, what matters to derive the error scaling behaviour analytically is our knowledge of $\overline{\Psi}$. This map depends on the initial state of the system and the way the controller $C$ interacts with the quantum system $S$.

\paragraph{Control error \& quantum information.--}

The transmission of quantum information over a quantum channel can be quantified by the quantum capacity $Q$ measuring the rate of information that can be reliably transmitted (thus, without any degradation) through the channel. More formally, given a set of $n$ arbitrary quantum information carriers, the quantum capacity $Q$ is defined as the maximum value of the ratio $\kappa/n$, where $\kappa$ denotes the number of qubits effectively employed (e.g., faithfully transmitted within a communication link) in the implemented operation\,\cite{ShorPRA1995,BennetTIT2002,GyongyosiCST2018}. It is worth noting that the formal derivation of $Q$ ideally stems from the asymptotic limit of $\kappa$ and $n$ infinitely large, namely by hypothetically considering unlimited resources. Therefore, this mathematical (upper bound) definition usually cannot be calculated and, indeed, analytical closed formula have been found only in few cases, though approximated values of $Q$ can be computed by means of numerical simulations or empirical analysis\,\cite{Caruso_RevModPhys_2014}.

We consider the complementary channel $\overline{\Psi}[\rho]$, mapping $\rho_{C}$ into $\widetilde{\rho}_{S}$, and the corresponding quantum capacity $Q_{\bar{\Psi}}$ that quantifies the maximum rate of information needed by $C$ to control the quantum system $S$. Thus, the performance in controlling $S$ necessarily depends on the value of $Q_{\bar{\Psi}}$, and here we investigate if there exists a formal relation expressing the error in controlling $S$ as a function of $Q_{\bar{\Psi}}$. To evaluate this aspect, let us consider the information-theoretic error bound proposed in Ref.\,\cite{LloydPRL2014}, which establishes how much a classical control action for a quantum system can be \emph{informative}. One can divide the space of possible target states $\hat{\rho}_S$ of $S$ into hyperspheres of radius $\varepsilon$ (called $\varepsilon$-balls), so that, if the final state $\widetilde{\rho}_{S}$ perfectly (i.e., without error) overlaps at least with one state in each $\varepsilon$-ball, then any other target state in the $\varepsilon$-ball can be reached with a control error $\leq\varepsilon$. As a consequence, the number of independent controls, which we have to be able to realize, has to correspond at least to the number of $\varepsilon$-balls. Note that we are implicitly taking into account also the possibility that the target state $\hat{\rho}_S$ is not reachable, namely that the equality $\widetilde{\rho}_{S} = \hat{\rho}_S$ cannot be achieved. The self-information associated to each $\varepsilon$-ball is equal to $-D\log_{2}(\varepsilon)$, where $D$ is the dimension of the state space. This implies that, regardless of how the control procedure is implemented, classically (modulation of the system Hamiltonian via an external classical control pulse) or quantum-mechanically (see Fig.\,\ref{fig:fig1}), the information content $I_c$ of the control action has to be at least greater or equal to the information associated to the $\varepsilon$-ball:
\begin{equation}\label{eq:infor_content}
I_{c}  \geq - D\log_{2}(\varepsilon).
\end{equation}
It follows that Eq.\,(\ref{eq:infor_content}) can be interpreted as the concept that a limited amount of information encoded in the control necessarily imposes a bound on the control error. In particular, from Eq.\,(\ref{eq:infor_content}) one finds that 
\begin{equation}
-\frac{I_c}{D} \leq \log_{2}(\varepsilon)\,\,\,
\Longleftrightarrow\,\,\,\varepsilon \geq 2^{-I_c/D}.
\end{equation}
According to the principles of information theory, which go back to Shannon's theorems\,\cite{Shannon48,Shannon49}, it has been established that the information content enclosed by a given (logic, computing, communication, control, etc) state is directly proportional to the product of two quantities: the \emph{bandwidth}, i.e., the maximum rate with which the information is transferred, and a logarithmic term that 
ideally tends to infinite when the accuracy in performing the desired operation on the state is maximum. By applying these concepts to the fully-quantum control problem described in Fig.\,\ref{fig:fig1}, the information content of a control action can be written as
\begin{equation}
I_c = n\,Q_{\bar{\Psi}}\log_{2}(1+\Delta r/\delta r)
\end{equation}
which can be interpreted as the quantum version of the Shannon-Hartley theorem\,\cite{CoverBook}, where $\Delta r$ and $\delta r$ are, respectively, the maximum range and the resolution of the control parameters entering in $\rho_C$\,\cite{footnote_delta_r}. The parameter $n$ is the number of repetitions of the transformation (with re-initialization of $\rho_C$) or the number of independent quantum controllers. As a result, the lower bound of the control error obeys the following relation:
\begin{equation}\label{eq:bound_error_q_control}
\log_{2}(\varepsilon) \geq -\frac{n\,Q_{\bar{\Psi}}}{D}\log_{2}\left(1+\frac{\Delta r}{\delta r}\right).
\end{equation}
It is worth noting that in real-world control problems the bound (\ref{eq:bound_error_q_control}) can largely depend on the ratio $\Delta r/\delta r$ -- the accuracy with which $\rho_C$ is sampled -- that in turn relies on the metric chosen for the parametrization of the control space. For this reason, the bound needs to be calibrated, and thus we introduce the parameters $a_1$ and $a_2$ with the aim to obtain an unique scaling behaviour of $\varepsilon$ as a function of the quantum capacity $Q_{\bar{\Psi}}$, i.e.,
\begin{equation}\label{eq:bound_error_q_control_2}
    \log_2(\varepsilon)\geq -a_1 Q_{\bar{\Psi}} - a_2 \ .
\end{equation}
Below, some analytical examples, involving a discrete and continuous variable quantum system, are presented to show the effectiveness of the bound\,(\ref{eq:bound_error_q_control_2}) for the scaling of $\varepsilon$.

\paragraph{Qubit-qubit control scheme.--}

As first example, let us consider that both the system $S$ and the controller $C$ are qubits\,\cite{footnote_qubits}. In this context, the reduced dynamics of $C$ induced by system-controller interactions can be described as a map $\Psi$ represented by only two Kraus operators\,\cite{Caruso_RevModPhys_2014}:
\begin{equation*}\label{eq:qubit_param}
A_{1}=\begin{pmatrix} \cos\theta & 0 \\ 0 & \cos\varphi \end{pmatrix}
\,\,\,\text{and}\,\,\,
A_{2}=\begin{pmatrix} 0 & \sin\varphi \\ \sin\theta & 0 \end{pmatrix}
\end{equation*}
such that $\Psi[\rho_{C}] \equiv \widetilde{\rho}_C = A_{1}\rho_{C}A_{1}^{\dagger} + A_{2}\rho_{C}A_{2}^{\dagger}$. This parametrization describes a wide class of two-qubit interactions as, e.g., the amplitude damping channel for $\cos(2\theta)=1$, $\cos(2\varphi)=2\eta - 1$ (with $\eta$ damping rate), or the bit-flip channel when $\theta=\varphi$. In particular, as proved in \cite{GiovannettiPRA2005,CarusoPRA2007,WolfPRA2007}, if $\cos(2\theta)/\cos(2\varphi) < 0$ then $Q_{\Psi}=0$ and the quantum channel $\Psi$ is denoted as anti-degradable. Conversely, if $\cos(2\theta)/\cos(2\varphi) \geq 0$, $Q_{\Psi}$ is obtained by solving the optimization problem expressed in terms of the single-letter formula
\begin{equation*}
Q_{\Psi} = \max_{p\in[0,1]}\mathcal{S}(c_1) - \mathcal{S}(c_2) 
\end{equation*}
with $p$ a real number, $\mathcal{S}(x) \equiv -x\log_{2}(x) - (1-x)\log_{2}(1-x)$ the binary Shannon entropy function, and $c_1 = p\cos^{2}(\theta) + (1-p)\sin^{2}(\varphi)$, $c_2 = p\sin^{2}(\theta) + (1-p)\sin^{2}(\varphi)$. Moreover, the quantum capacity $Q_{\bar{\Psi}}$ of the complementary channel $\overline{\Psi}$ can be directly derived from $Q_{\Psi}$ by means of the relation
\begin{equation*}
Q_{\bar{\Psi}} = Q_{\Psi}\left(\theta \rightarrow -\overline{\theta}, \varphi \rightarrow \overline{\varphi} + \frac{\pi}{2}\right)
\end{equation*}
where $\widetilde{\rho}_{S} \equiv \overline{\Psi}[\rho_C] = \overline{A}_{1}\rho_{C}\overline{A}_{1}^{\dagger}+\overline{A}_{2}\rho_{C}\overline{A}_{2}^{\dagger}$ and $\overline{A}_{k} \equiv A_{k}(\theta \rightarrow -\overline{\theta}, \varphi \rightarrow \overline{\varphi} + \frac{\pi}{2})$ with $k=1,2$.

\begin{figure}[t!]
\centering
\includegraphics[scale=1.3]{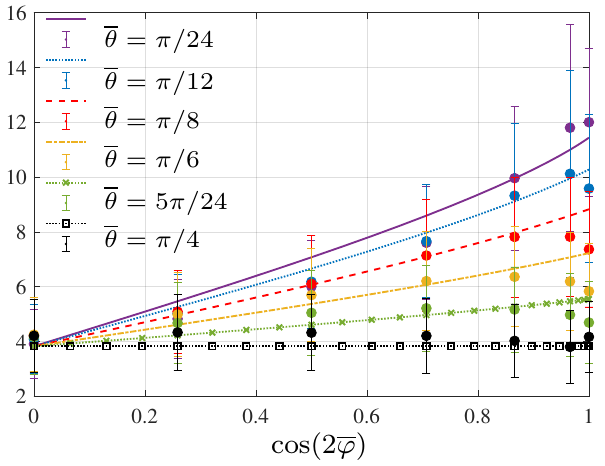}
\caption{\emph{Qubit-qubit control scheme}.
Comparison between the theoretical bound $a_{1}Q_{\bar{\Psi}} + a_{2}$ (lines), with $Q_{\bar{\Psi}}\in[0,1]$, and $-\log_{2}\langle\varepsilon\rangle$ (dots) as a function of $\cos(2\overline{\varphi})$ and 6 different values of $\cos(2\overline{\theta})$, with $\overline{\varphi}$, $\overline{\theta}=k\pi/24$ and $k=1,\ldots,6$. The error bars denote the standard deviation of the negative logarithm of the control error, while the values of the model parameters $a_1$ and $a_2$, respectively equal to $11.8$ and $3.8$, are obtained by means of a single fitting procedure operating at once on all the 6 curves depicted in the figure. The control error is obtained by solving numerically the proposed fully-quantum control problem for $1000$ different target states $\hat{\rho}_S$ and then making the average over all the sampled target states. It is worth noting that for $\cos(2\overline{\varphi})\rightarrow 1$ with $\overline{\theta}\neq 0$ the numerical average control error slightly increases (i.e. $-\log_{2}\langle\varepsilon\rangle$ slightly decreases when $\cos(2\overline{\varphi})$ is very close to $1$) -- see Appendix for more details.}
\label{fig:fig2}
\end{figure}
According to the fully-quantum control problem as defined in the previous section we have to find the optimal control state $\rho_C = \rho_{C}^{\star}$ such that the cost function (i.e., the control error) $\varepsilon = 1 - \mathfrak{F}\left(\hat{\rho}_{S},\overline{\Psi}[\rho_C]\right) \geq 0$ is minimized. Formally, for this example we can always find an analytical solution $\rho_{C}^{\star}$ such that $\widetilde{\rho}_S = \hat{\rho}_S$ (see Appendix). However, only if the target state $\hat{\rho}_S$ is reachable by the system does the formal solution $\rho_{C}^{\star}$ correspond to a physical state. In such case, the quantum system $S$ can be brought to the target state $\hat{\rho}_{S}$ with zero error. 

In Fig.\,\ref{fig:fig2} the information-theoretical error bound (lines) is compared with the average control error $\langle\varepsilon\rangle$ (dots) obtained from numerical simulations, both as a function of $\cos(2\overline{\theta})$ and $\cos(2\overline{\varphi})\in[0,1]$. For each dot plotted in Fig.\,\ref{fig:fig2}, the average control error is computed over $1000$ random target states $\hat{\rho}_S$, uniformly sampled from all the Bloch sphere by respecting the Haar measure\,\cite{footnote_state_sampling}. The information-theoretical bound for $\varepsilon$ is given by Eq.~(\ref{eq:bound_error_q_control_2}), where the values of the parameters $a_1$ and $a_2$ are determined by means of a least-squares fitting procedure. Also the maximum values $\varepsilon_{\rm max}$ of the control error (i.e., the respective maximum over the 1000 random target state for each set of parameters) have been analyzed: apart from a scale factor, namely slightly different numbers of $a_1$ and $a_2$, their behaviour is qualitatively comparable with the one obtained for the average values (see figure in the Appendix). The agreement between theory and numerical simulations is very good. As discussed in the Appendix, we have tested the scaling behaviour of the logarithm of the control error also by using fitting models with more than $2$ free-parameters, i.e., $-b_{1}Q_{\bar{\Psi}}^{b_{3}} - b_2$ and $-c_{1}\log_{2}(Q_{\bar{\Psi}} + c_3)-c_{2}$. Overall, our analysis confirms that the average control error scales as a power of $2$ proportionally to $Q_{\bar{\Psi}}$, as described by Eq.\,(\ref{eq:bound_error_q_control_2}). This leads us to conclude that, apart from a few single parameter values (e.g., the limit case of $\cos(2\overline{\varphi}) \rightarrow 1$ with $\overline{\theta} \neq 0$ discussed in the Appendix), the error associated to the fully-quantum control procedure follows the information-theoretical model and tends to zero when the value of the quantum capacity $Q_{\bar{\Psi}}$ is maximized.

\paragraph{One-mode bosonic Gaussian channels.--}

As second and more challenging example, $S$ and $C$ are taken as continuous-variable systems described in terms of one-mode bosonic harmonic oscillators, typically a specific normal mode of the electromagnetic field. In particular, we consider the so-called Gaussian quantum channels mapping Gaussian (i.e. with a Gaussian characteristic function) input states to Gaussian output states\,\cite{HolevoPRA2001} and it is experimentally widespread, since it includes not only linear attenuation and amplification processes, but also thermalization and squeezing phenomena and any physical interaction described by a quadratic Hamiltonian. As discussed in \cite{FerraroReview2005,Caruso_RevModPhys_2014}, they can be described in terms of a single parameter $K \geq 0$ and are unitarily equivalent to two (canonical) classes of Gaussian channels, simply corresponding to attenuation and amplification processes respectively. For $K^2 \leq 1$, the quantum channel corresponding to the process describes linear losses with attenuation factor $K^2$, while for $K^2 > 1$ an amplification with gain $K^2$ is obtained\,\cite{FerraroReview2005,Caruso_RevModPhys_2014}. Moreover, at the level of quantum capacity, $Q_{\Psi}=0$ for $K^2 \leq 1/2$; in such a case the channel is called anti-degradable. Otherwise, when $K^2 > 1/2$, $Q_{\Psi}=\log_{2}(K^2/|K^2 - 1|)$ leading to the degradable channel case\,\cite{CarusoNJP2006}. For our control purposes, also in this example the quantum capacity $Q_{\bar{\Psi}}$ associated to the complementary (still Gaussian) channel $\overline{\Psi}$ is determined straightforwardly from the knowledge of $Q_{\Psi}$, as in the qubit-qubit control scheme previously analyzed. Specifically, by means of the functional substitution $K^2 \rightarrow 1 - \overline{K}^2$, one gets $Q_{\bar{\Psi}}=\log_{2}(|1 - \overline{K}^2|/\overline{K}^2)$ with $\overline{K}^2 \geq 0$. From here on, for the sake of simplicity of notation, we will denote $\overline{K}^2$ as $q$. 

\begin{figure}[t!]
\centering
\includegraphics[scale=0.52]{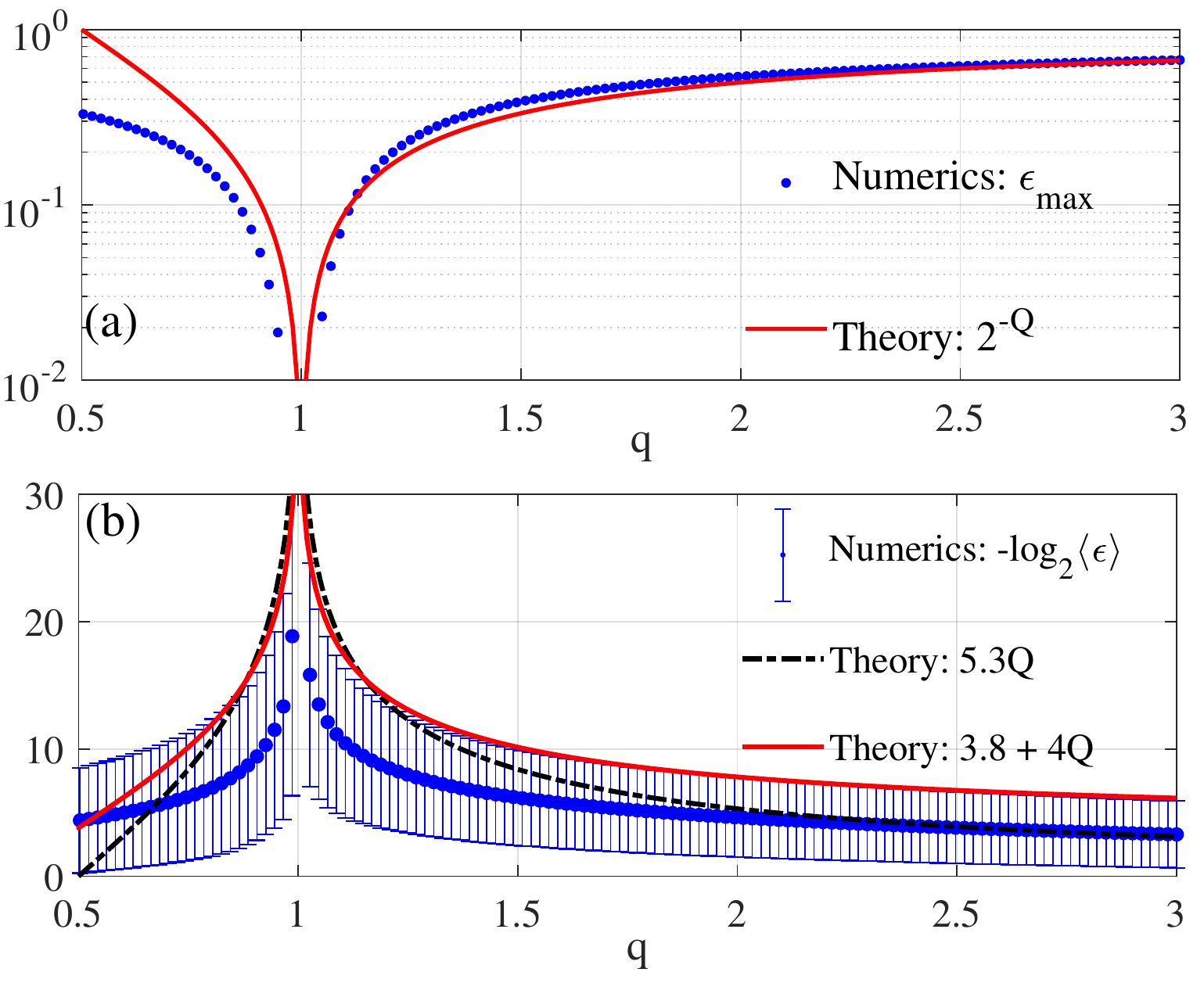}
\caption{\emph{Fully-quantum control associated to a one-mode Bosonic Gaussian channel}. 
(a) Comparison between the maximum values $\varepsilon_{\rm max}$ of the control error $\varepsilon$ (blue dots), obtained by numerically solving the control problem for $1000$ different target covariance matrices, and $2^{-Q_{\bar{\Psi}}} \equiv q/|1-q|$ (red solid line) for $q\in[0.5,3]$ by using a logarithmic scale for the $y$-axis (for the sake of brevity, $Q_{\bar{\Psi}}$ is denoted as $Q$ in the legends of the two panels of the figure). Note that, although the comparison in (a) is more qualitative (indeed, $a_1=a_2=1$), the link between $\varepsilon_{\rm max}$ and $Q_{\bar{\Psi}}$ can be clearly assessed. (b) Linear-scale comparison between the numerical values of $-\log_{2}\langle\varepsilon\rangle$ and the empirical model $a_{1}Q_{\bar{\Psi}} + a_{2}$ as a function of $q\in[0.5,3]$. The blue dots denote the average values $\langle\varepsilon\rangle$ of the control error over the sampled target covariance matrices, while the corresponding standard deviations are represented by blue error bars. Here, the values of $a_1$ and $a_2$ are respectively provided by the sets $(5.3,0)$ (black dotted line) and $(4,3.8)$ (red solid line). Also notice that $Q_{\bar{\Psi}}\in[0,\infty)$: $Q_{\bar{\Psi}}=0$ for $q = 1/2$ and $q \rightarrow \infty$, while $Q_{\bar{\Psi}}\rightarrow\infty$ when $q = 1$.}
\label{fig:fig3}
\end{figure}
To control the Gaussian state of the quantum system $S$, we need to search for the covariance matrix $\gamma_C$, associated to the control Gaussian state $\rho_C$, ensuring that $\widetilde{\gamma}_S \equiv X^{T}\gamma_{C}X + Y = \hat{\gamma}_S$, where $\hat{\gamma}_S$ is the target covariance matrix and $\widetilde{\gamma}_S$ denotes the covariance matrix of the final (Gaussian) state $\widetilde{\rho}_S$ of the system $S$. Note that, in doing this, we are implicitly assuming (without loss of generality) that the quantum channels $\Phi$ and $\Psi$ are (unitarily) reduced to a canonical form, i.e. with vanishing displacements and with the matrices $X$ and $Y$ taking a particular symmetric form \cite{SerafiniPRA2005,Caruso_RevModPhys_2014}. Hence, $X \equiv \sqrt{q}\mathbbm{1}$ and $Y \equiv |q - 1|\mathbbm{1}$, where $\mathbbm{1}$ denotes the identity matrix, and the solution to the control problem can be analytically determined. In particular, the optimal control covariance matrix $\gamma_{C}^{\star}$, allowing for $\widetilde{\gamma}_S = \hat{\gamma}_S$ with zero error, is equal to
\begin{equation}\label{eq:optimal_covariance}
\gamma_{C}^{\star} = \begin{pmatrix}
\frac{\hat{\gamma}_{1}-|q - 1|}{q} & \frac{\hat{\gamma}_{2}}{q} \\
\frac{\hat{\gamma}_{2}}{q} & \frac{\hat{\gamma}_{3}-|q - 1|}{q}
\end{pmatrix}\,\,\,\,\,\text{with}\,\,\,\,\,\,\hat{\gamma}_{S} \equiv \begin{pmatrix} \hat{\gamma}_{1} & \hat{\gamma}_{2} \\ \hat{\gamma}_{2} & \hat{\gamma}_{3} \end{pmatrix},
\end{equation}
provided that $\gamma_{C}^{\star}$ obeys the generalized uncertainty relation $\gamma_{C}^{\star} \geq i\sigma$, where $\sigma$ is the single-mode phase-space canonical symplectic matrix (taking into account all the commutation relations of the the ladder operators). If this inequality holds, then $\gamma_{C}^{\star}$ corresponds to a physical state. Otherwise, the optimal control covariance matrix is chosen as the covariance matrix that minimizes the control error $\varepsilon = 1 - \mathfrak{F}\left(\hat{\gamma}_{S},\widetilde{\gamma}_S\right) \geq 0$, where $\mathfrak{F}\left(\hat{\gamma}_{S},\widetilde{\gamma}_S\right) \equiv {\rm Tr}\sqrt{\sqrt{\hat{\gamma}_{S}}\widetilde{\gamma}_S\sqrt{\hat{\gamma}_{S}}}$ is the Uhlmann fidelity between the target and final covariance matrices, $\hat{\gamma}_{S}$ and $\widetilde{\gamma}_S$, respectively.

In Fig.\,\ref{fig:fig3} we compare for $q\in[0.5,3]$ the information-theoretical error bound (\ref{eq:bound_error_q_control}) with the the control error obtained by numerically solving the control problem for $1000$ different target covariance matrices, uniformly sampled from the space of one-mode Gaussian quantum states in accordance with the Haar measure defined on this space. In particular, the numerical findings are compared with the theoretical predictions provided by the quantum capacity $Q_{\bar{\Psi}}$, the analytic curve $2^{-Q_{\bar{\Psi}}}$, and the calibrated model of Eq.\,(\ref{eq:bound_error_q_control_2}). Also in this case, the models $-b_{1}Q_{\bar{\Psi}}^{b_{3}} - b_2$ and $-c_{1}\log_{2}(Q_{\bar{\Psi}} + c_3)-c_{2}$, defined by more than $2$ free-parameters, are tested. Similarly to the qubit-qubit control scheme, the values of the model parameters are chosen by means of a least-squares fitting procedure. We have found that the correct scaling of both the average control error $\langle\varepsilon\rangle$ (and corresponding confidence intervals defined by the error bars in Fig.\,\ref{fig:fig3}) and the maximum values $\varepsilon_{\rm max}$ is reproduced by the bound given by Eq.~(\ref{eq:bound_error_q_control_2}). The quantitative analysis of the fit, and corresponding error values, of the fitted models is presented in the Appendix. In conclusion, Fig.\,\ref{fig:fig3} confirms the main result discussed in this Letter, namely that for a quantum system (in this case, a continuous-variable one) the average control error, resulting by applying a fully-quantum control procedure, scales exponentially as $2^{-Q_{\bar{\Psi}}}$ with the associated quantum channel capacity. Thus, the system can be potentially controlled with zero error if the capacity of the complementary channel $\overline{\Psi}$ takes its maximum value.

\paragraph{Applications.--}

Here we identify three explicit applications for our results: (i) To perform \emph{quantum state preparation} of (many-body) quantum systems that are difficult both to access and control via classical control fields. In such cases one could use another quantum system with more control knobs as a quantum controller that allows for full control over the main system. This, for instance, may be experimentally implemented in state-of-the-art solid-state platforms exploiting nuclear spins as controller of large quantum registers of electron spins in diamond\,\cite{Taminiau14,Taminiau19}. (ii) To realize a \emph{photonic quantum bus}\,\cite{SimonNatPhys2007} being able to connect quantum memories and $n$-qubits quantum processors. Indeed, one could think to prepare single atoms or atomic ensemble in distinct (remote) cavities -- as done e.g.\,in\,\cite{Julsgaard2013,BarontiniScience2015,Morello2020} -- and link the atoms through single photons or few-photon quantum states. The control problem addressed here is the same as the one depicted in Fig.\,\ref{fig:fig1}, where the atoms in the cavities represent the system $S$ to be controlled and the photons are the quantum controller $C$. (iii) To carry out \emph{long-distance quantum communication} through flying photons\,\cite{HammererRMP10} by implementing quantum repeater protocols, as e.g.\,the one proposed by Duan-Lukin-Cirac-Zoller (DLCZ)\,\cite{DuanNature01}. Similarly to the previous case, pairs of entangled photons sent through the communication channel represent the auxiliary control systems, while the platforms used to realize the quantum memories of the scheme are the system that we aim to control. 

\paragraph{Conclusions.--}

In this Letter we have analytically characterised the scaling of the error $\varepsilon$ in controlling a quantum system $S$ through the interaction with an auxiliary one, i.e., the quantum controller $C$. Specifically, we have demonstrated that $\varepsilon$ scales as $2^{-Q_{\bar{\Psi}}}$ where $Q_{\bar{\Psi}}$ is the quantum capacity of the complementary channel $\overline{\Psi}$ mapping $\rho_C$ onto $\widetilde{\rho}_S$. Our theoretical findings are confirmed by numerical simulations (Figs.\,\ref{fig:fig2} and \,\ref{fig:fig3}) taking into account both discrete- and continuous-variable systems.

In all cases where the fully-quantum control procedure is required, one can take the quantum controller as a quantum system with the same dimension of the controlled one and, then, optimize the control parameters with $n=1$. Conversely, a lower-dimension controller could be employed, but one would need to choose $n>1$, i.e., to repeat the control operation -- with re-initialization of $\rho_C$ -- more than once. In any case, from the information-theoretic error scaling (\ref{eq:bound_error_q_control}) or (\ref{eq:bound_error_q_control_2}), we can deduce that the control over a system via a quantum controller is maximized (at given quantum controller and interaction between system and controller) if the initial state of the system is chosen such that the quantum channel capacity $Q_{\bar{\Psi}}$ is maximized. This generally allows for the best possible control at the lowest possible repetitions $n$ of the control operation. The discussed applications clearly show the very promising impact of the achieved results on many different fields involving quantum technologies.

\paragraph{Acknowledgments.--}
S.G., M.M.M. and F.C.\,acknowledge funding from the Fondazione CR Firenze through the project Q-BIOSCAN. S.G.\,and F.C.\,were financially supported from by the Fondazione CR Firenze through the project QUANTUM-AI, the European Union’s Horizon 2020 research and innovation programme under FET-OPEN Grant Agreement No.\,828946 (PATHOS), and from University of Florence through the project Q-CODYCES. M.M.M. and T.C. acknowledge funding from the European Union’s Horizon 2020 research and innovation programme under Grant Agreement No.\,817482 (PASQuanS), as well as from the Deutsche Forschungsgemeinschaft (DFG, German Research Foundation) under Germany’s Excellence Strategy – Cluster of Excellence Matter and Light for Quantum Computing (ML4Q) EXC 2004/1 – 390534769. S.M. acknowledges support from the European Union’s Horizon 2020 research and innovation programme under the Marie Skłodowska-Curie grant agreement No. 765267 (QuSCo), and No. 817482 (PASQuanS), by the Italian PRIN 2017 and the CARIPARO project QUASAR.

\section{Appendix}

\subsection{Control error curve fitting}

Here, we provide further details on the numerical simulations of Figs.\,\ref{fig:fig2} and \ref{fig:fig3}. In particular, both for the qubit-qubit control scheme and the control procedure using one-mode bosonic Gaussian channels, we will show the results obtained by numerically testing three different models for the control error scaling. In this regard, notice that for both cases the control error $\varepsilon$ is obtained by solving the fully-quantum control procedure described in the main text. Given the control error $\varepsilon$, the models that we have tested (below denoted as $\mathcal{M}$) by making the comparison with $-\log_{2}(\varepsilon)$ are the following: 
\begin{enumerate}[(i)]
    \item 
    $\mathcal{M} = a_{1}Q_{\bar{\Psi}} + a_2$
    \item
    $\mathcal{M} = \displaystyle{b_{1}Q_{\bar{\Psi}}^{b_{3}}} + b_2$
    \item
    $\mathcal{M} = c_{1}\log_{2}(Q_{\bar{\Psi}} + c_3)+c_{2}$\,.
\end{enumerate}
In turn, it is worth observing that the models (i)-(iii) correspond to the following models $M$ for the scaling behaviour of $\varepsilon$: 
\begin{enumerate}[(I)]
\item
$M = a_{3}\,2^{-a_{1}Q_{\bar{\Psi}}}$ with $a_3 = 2^{-a_2}$
\item
$M = b_{4}\,2^{-b_{1}Q_{\bar{\Psi}}^{b_3}}$ with $b_4 = 2^{-b_2}$
\item
$M = c_{4}(Q_{\bar{\Psi}} + c_3)^{-c_1}$ with $c_4 = 2^{-c_2}$.
\end{enumerate}

Below, we will show that the scaling given by (i) (or equivalently (I)) is the best solution in terms of the fitting error $\zeta(\varepsilon)$ and/or the number of parameters adopted for the fitting. The fitting error $\zeta(\varepsilon)$ is defined as the ratio between the Euclidean distance (or $L^2$ norm) of the difference between $-\log_{2}(\varepsilon)$ and the corresponding fitting model, and the Euclidean distance $-\log_{2}(\varepsilon)$ alone. More formally,
\begin{equation*}
  \zeta(\varepsilon) \equiv \frac{\|\mathcal{M} +\log_{2}(\varepsilon)\|_2}{\|\log_{2}(\varepsilon)\|_2}\,.
\end{equation*}

\paragraph{Qubit-qubit control scheme.--}

For the example with the qubit-qubit control scheme, the fully-quantum control procedure requires to find the optimal value $y^{\star}$ and $z^{\star}$ of the parameters $y$ (real number) and $z$ (complex number) pertaining to the control state
\begin{equation*}
   \rho_C = \begin{pmatrix} y & z \\ z^{\ast} & 1-y \end{pmatrix} 
\end{equation*}
such that $\widetilde{\rho}_{S}$, the final density operator of $S$ after the control transformation, is as close as possible to the target state 
\begin{equation*}
\hat{\rho}_S \equiv \begin{pmatrix} \hat{y} & \hat{z} \\ \hat{z}^{\ast} & 1-\hat{y} \end{pmatrix}.
\end{equation*}
Being $\widetilde{\rho}_{S}=\overline{A}_{1}\rho_{C}\overline{A}_{1}^{\dagger}+\overline{A}_{2}\rho_{C}\overline{A}_{2}^{\dagger}$ with $\overline{A}_{1}$ and $\overline{A}_{2}$ the quantum maps associated to the complementary channel $\overline{\Psi}$, one can proceed to solve the equation
$$
\hat{\rho}_{S}=\overline{A}_{1}\rho_{C}^{\star}\overline{A}_{1}^{\dagger}+\overline{A}_{2}\rho_{C}^{\star}\overline{A}_{2}^{\dagger}
$$
as a function of the elements of the optimal control state $\rho_{C}^{\star}$ and find the analytical expressions of $y^{\star}$ and $z^{\star}$. The latter are given by the following relations:
\begin{eqnarray*}
&y^{\star} = \displaystyle{\frac{\hat{y}-\sin^{2}(\overline{\varphi})}{\cos^{2}(\overline{\theta})-\sin^{2}(\overline{\varphi})}}&\nonumber \\
&\displaystyle{\mathfrak{Re}\{z^{\star}\} = \frac{-\mathfrak{Re}\{\hat{z}\}}{\cos(\overline{\varphi}-\overline{\theta})}}\,\,\,\,\,\text{and}\,\,\,\,\,\displaystyle{\mathfrak{Im}\{z^{\star}\} =
\frac{-\mathfrak{Im}\{\hat{z}\}}{\cos(\overline{\varphi}+\overline{\theta})}}&
\end{eqnarray*}
where $\mathfrak{Re}\{x\}$ and $\mathfrak{Im}\{x\}$ denote the real and imaginary part of the generic complex number $x$, respectively. However, not always the obtained solutions are physically feasible. In particular, if $y^{\star}(1-y^{\star})-|z^{\star}|^2 \geq 0$, then the optimal control state $\rho_{C}^{\star}$ is physically realizable and one can get the equality $\widetilde{\rho}_{S}=\hat{\rho}_{S}$ with zero error. Otherwise, the optimal control state $\rho_{C}^{\star}$ is obtained as the (physical) state minimizing the cost function $\varepsilon = 1 - \mathfrak{F}\left(\hat{\rho}_{S},\widetilde{\rho}_S\right) \geq 0$, with $\mathfrak{F}(\hat{\rho}_{S},\widetilde{\rho}_S) \equiv {\rm Tr}\sqrt{\sqrt{\hat{\rho}_{S}}\widetilde{\rho}_S\sqrt{\hat{\rho}_{S}}}$ Uhlmann fidelity. The latter is the procedure that has been followed to derive the control error $\varepsilon$ in the numerical simulations. Specifically, $\varepsilon$ has been computed as a function of $\cos(2\overline{\theta})$ and $\cos(2\overline{\varphi})$, both belonging to the interval $[0,1]$, for $1000$ random final target states $\hat{\rho}_S$ uniformly sampled from all the Bloch sphere by respecting the Haar measure. The negative binary logarithm of the average value of $\varepsilon$, i.e.\,$-\log_{2}\langle\varepsilon\rangle$, has been compared with the models (i)-(iii), all originating by the information-theoretic error bound of Eq.\,(\ref{eq:bound_error_q_control}) in the main text. For all the models we now provide the values of the set of parameters $\{a_k\}_{k=1}^{2}$, $\{b_k\}_{k=1}^{3}$ and $\{c_k\}_{k=1}^{3}$, obtained by means of a least-squares fitting procedure, and the corresponding error values $\zeta(\langle\varepsilon\rangle)$, i.e.,
\begin{enumerate}[(i)]
    \item 
    $a_{1}=11.8$,\,\,\,$a_{2}=3.8$;\,\,\,$\zeta=0.086$
    \item
    $b_{1}=14.1$,\,\,\,$b_{2}=4.3$,\,\,\,$b_{3}=1.4$;\,\,\,$\zeta=0.064$
    \item
    $c_{1}=11.3$,\,\,\,$c_{2}\approx 0$,\,\,\,$c_{3}=1.27$;\,\,\,$\zeta=0.094$\,.
\end{enumerate}
By analysing only the error values $\zeta(\langle\varepsilon\rangle)$ (all smaller than $0.1$) obtained by the fitting procedure, one can deduce that the best result is given by model (ii). However, all the error values $\zeta(\langle\varepsilon\rangle)$ are very close to each other. Thus one can reliably state that the results from model (i) are consistent with the ones from models (ii) and (iii). Moreover, also by comparing the behaviour of $-\log_{2}\langle\varepsilon\rangle$ as a function of $\cos(2\overline{\varphi})$ and $\cos(2\overline{\theta})$, the three models can be considered as equivalent within the relevant interval (the values assumed by $Q_{\bar{\Psi}}$). To determine the choice of the most suitable model for the control error scaling, we resort to minimal complexity arguments, whereby the model to be privileged is the one with the lower number of free-parameters/coefficients and the same fitting error. Thus, our choice falls on model (i) that just uses two free-parameters. This confirms that the average control error scales exponentially with the negative quantum capacity $Q_{\bar{\Psi}}$. 
\begin{figure}[t!]
\centering
\includegraphics[scale=1.32]{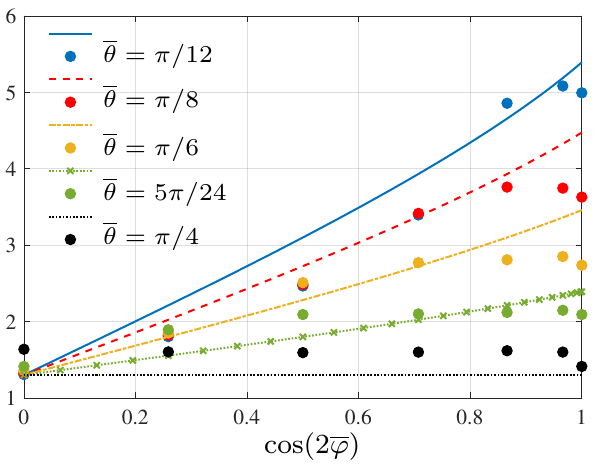}
\caption{Comparison between the theoretical bound $a_{1}Q_{\bar{\Psi}} + a_{2}$ (lines), with $Q_{\bar{\Psi}}\in[0,1]$, and $-\log_{2}\varepsilon_{\rm max}$ (dots) as a function of $\cos(2\overline{\varphi})$ and 5 different values of $\cos(2\overline{\theta})$ corresponding to $\overline{\theta}=k\pi/24$ with $k=2,\ldots,6$. Again the values of the model parameters $a_1$ and $a_2$, here respectively equal to $7.5$ and $1.3$, are obtained by means of a single fitting procedure operating at once on all the 5 curves depicted in the figure.}
\label{fig:fig_appendix}
\end{figure}
Furthermore, it is worth recalling that also the maximum values $\varepsilon_{\rm max}$ of the control error have been analyzed. Also in this case, as shown in Fig.\,\ref{fig:fig_appendix}, a scaling according to model (i) comparable with the one obtained for the average values, can be observed. 

Now, let us discuss more in detail the aspects regarding the increasing of the average control error $\langle\varepsilon\rangle$ (namely, the decreasing of $-\log_{2}\langle\varepsilon\rangle$) for $\cos(2\overline{\varphi})=1$ and $\overline{\theta} \neq 0$ or $\cos(2\overline{\theta})=1$ and $\overline{\varphi} \neq 0$. In doing this, let us analyze the Kraus operators $\overline{A}_{1}$ and $\overline{A}_{2}$ that are involved in the control procedure. Such operators era equal respectively to
\begin{equation*}\label{eq:qubits_parametrization}
\overline{A}_{1}=\begin{pmatrix} \cos\overline{\theta} & 0 \\ 0 & -\cos\overline{\varphi} \end{pmatrix}
\,\,\,\text{and}\,\,\,
\overline{A}_{2}=\begin{pmatrix} 0 & \sin\overline{\varphi} \\ -\sin\overline{\theta} & 0 \end{pmatrix}
\end{equation*}
with $\overline{\theta}$, $\overline{\varphi}\in[0,\frac{\pi}{4}]$ so as to ensure that $Q_{\bar{\Psi}}>0$. In particular, when $\cos(2\overline{\varphi})=1$ and $\overline{\theta} \neq 0$ or $\cos(2\overline{\theta})=1$ and $\overline{\varphi} \neq 0$, $\overline{A}_{2}$ becomes a singular operator and both of its eigenvalues are equal to zero. This means that, in such a case, the operator $\overline{A}_{2}$ is nilpotent. The singularity of the Kraus operator is the reason under the slight rising of the control error values, which in turn can be interpreted as a reduction of the dimension of the space of control states. Finally, it is also worth noting that the simultaneous validity of the conditions $\cos(2\overline{\varphi})=1$ and $\cos(2\overline{\theta})=1$, i.e.\,$\overline{\varphi}=\overline{\theta}=0$, is not pathological in the sense that, apart from a phase factor, the solution to the control problem is just provided by the equality $\rho_{C}^{\star}=\hat{\rho}_{S}$.

\paragraph{Control procedure with one-mode Bosonic Gaussian channels.--}

Any Gaussian quantum state is fully characterized by its first and second moments (of the characteristic function in the phase-space representation), also denoted as \emph{displacement vector} and \emph{covariance matrix}, respectively. For the fully-quantum control procedure depicted in Fig.\,\ref{fig:fig1} in the main text, we assume that the quantum channels governing the reduced dynamics of $S$ and $C$ are described as Gaussian channels, mapping Gaussian states into Gaussian ones. By applying suitable Gaussian unitaries at the input and output of the channel, one can always neglect the first moment contributions and exploit a particular symmetric form for the matrices $X$ and $Y$, hence obtaining the canonical form of the channel in terms of evolution of the covariance matrix of the considered system \cite{Caruso_RevModPhys_2014}. Therefore, in our case we have to look for the covariance matrix $\gamma_C$, related to the control state $\rho_C$, such that 
\begin{equation*}
\widetilde{\gamma}_S \equiv X^{T}\gamma_{C}X + Y = \hat{\gamma}_S
\end{equation*}
with $X \equiv \sqrt{q}\mathbbm{1}$, $Y \equiv |q - 1|\mathbbm{1}$, $q$ a real number greater than $1/2$, and $\hat{\gamma}_S$ being the target covariance matrix for the system state. Then, given the optimal covariance matrix $\gamma_{C}^{\star}$ provided by Eq.\,(\ref{eq:optimal_covariance}) in the main text, if the generalized uncertainty relation  $\gamma_{C}^{\star} \geq i\sigma$ (with $\sigma$ being the canonical symplectic matrix) holds, then the controller state physically exists and the control task can be carried out with zero error. Otherwise, if $\gamma_{C}^{\star} < i\sigma$, the optimal control covariance matrix $\gamma_{C}$ is taken so as to minimize the cost function $\varepsilon = 1 - \mathfrak{F}\left(\hat{\gamma}_{S},\widetilde{\gamma}_S\right)$, with $\mathfrak{F}\left(\hat{\gamma}_{S},\widetilde{\gamma}_S\right) \equiv {\rm Tr}\sqrt{\sqrt{\hat{\gamma}_{S}}\widetilde{\gamma}_S\sqrt{\hat{\gamma}_{S}}}$, but while satisfying the uncertainty relation $\gamma_{C} \geq i\sigma$.

On the numerical side, the control error $\varepsilon$ is computed as a function of $q\in[0.5,3]$ and for $1000$ different target covariance matrix, uniformly sampled from the space of one-mode Gaussian quantum states in accordance with the Haar measure. Then, both the average and the maximum values of $\varepsilon$, $\langle\varepsilon\rangle$ and $\varepsilon_{\rm max}$ respectively, have been evaluated. As shown in Fig.\,\ref{fig:fig3}\,(a) in the main text, the agreement between the maximum control error $\varepsilon_{\rm max}$ and the bound $2^{-Q_{\bar{\Psi}}}$ is very good, especially for $q\in[1,3]$. On the other hand, regarding the negative binary logarithm of the average control error $\langle\varepsilon\rangle$, we made use of the models (i)-(iii), as previously done for the qubit-qubit control scheme. The following results have been found: 
\begin{enumerate}[(i)]
    \item 
    $a_{1}=5.3$,\,\,\,$a_{2}\approx 0$;\,\,\,$\zeta=0.36$
    \item
    $b_{1}=5.3$,\,\,\,$b_{2}\approx 0$,\,\,\,$b_{3}=0.64$;\,\,\,$\zeta=0.14$
    \item
    $c_{1}=5.16$,\,\,\,$c_{2} \approx 0$,\,\,\,$c_{3}=1.1$;\,\,\,$\zeta=0.27$\,.
\end{enumerate}
The fitting procedure has been carried out by taking into account all the values of $Q_{\bar{\Psi}}$ pertaining to $q\in[0.5,3]$. Instead, for each 
computed set of model parameters, the fitting error $\zeta$ just refer to the values of $-\log_{2}\langle\varepsilon\rangle$ within the interval $[5,15]$ so as to prevent that $\zeta$ is biased by too large or too small values of the logarithm function. Thus, by analysing the figure of merit $\zeta$, the best results are provided by model (ii), but the values of the fitting error $\zeta$ for the models (i)-(iii) are comparable and of the same order of magnitude. For this reason and for the scaling of $-\log_{2}\langle\varepsilon\rangle$ in the analysed intervals, the three models can be considered consistent and with only slight differences among them. However, among the three models, only model (i) is characterized by $2$ free-parameters/coefficients, differently to models (ii)-(iii) that are defined by $3$ coefficients. Therefore, by resorting again to minimal complexity arguments, we conclude that the preferable model for the control error is the one predicted by our theoretical analysis, namely the one provided by model (i) that has the lower number of coefficients.




\begin{thebibliography}{99}
\bibitem{footnote_delta_r}
The maximum range $\Delta r$ is provided by the maximum difference between the largest and smallest values of the control parameters in $\rho_C$, while $\delta r$ denotes their resolution, i.e., the smallest interval allowing for two distinct values with different effects on the control action.
\bibitem{footnote_qubits}
Here, the idea is to try reducing as much as possible the control complexity, so as to fulfill the control task by taking $n=1$ and $C$ as a qubit. In this way, the control parameters are just given by one of the elements on the diagonal of $\rho_C$ and the real and imaginary parts of the corresponding coherence term.
\bibitem{footnote_state_sampling}
Note that the initial state of $S$ is already contained in the parameters of the Kraus maps $A_1$ and $A_2$, as well as of $\overline{A}_1$ and $\overline{A}_2$. Thus, being the initial states fixed by the map, in our numerical simulations we need to randomly choose just the target states.
\end{thebibliography}
\end{document}